\documentclass[11pt, notitlepage, reqno]{amsart}% Document starts from the title page and eqs numbers appear on the right
\usepackage[utf8]{inputenc}
\usepackage[T1]{fontenc}
\usepackage[british]{babel}

\usepackage{times}
\usepackage{amsaddr, amsbsy, amscd, amsfonts, amsmath, amssymb, amsthm, dsfont, mathrsfs, mathtools}
\usepackage[pdfusetitle,
        pdfauthor={G Basti, D Ferretti, A Teta},
		pdfsubject={Paper in Mathematical Physics},%
		pdfkeywords={%
                Zero-Range Interactions, Efimov Effect, Three-Body Hamiltonians%
            }]{hyperref}
\usepackage{enumitem}
\usepackage{xparse}
\usepackage{soul, xcolor}
\usepackage[bottom, hang, flushmargin]{footmisc}
\usepackage{eqnarray}
\usepackage{scalerel, stackengine}
\usepackage{fancyhdr}
\usepackage{indentfirst}
\usepackage[square,comma,numbers]{natbib}
\usepackage[a4paper, hmargin=2.5cm, vmargin={1.5cm,2.5cm}, nomarginpar, hcentering, includeheadfoot]{geometry}
\usepackage[toc, page]{appendix}
\let\oldtocsection=\tocsection
\let\oldtocsubsection=\tocsubsection

\renewcommand{\tocsection}[2]{%
    \bfseries\oldtocsection{#1}{#2}%
}
\renewcommand{\tocsubsection}[2]{%
  \hspace{1.5em}%
  \oldtocsubsection{#1}{#2}
}
\makeatletter
\renewcommand{\section}{%
    \@startsection{section}%
    {1}% level
    {0em}% indent
    {1.5cm \@plus 0.1ex \@minus -0.05ex}% Beforeskip
    {0.75cm \@plus 0.2em}% Afterskip
    {\centering \Large \scshape}% Style
}

\renewcommand{\subsection}{%
    \@startsection{subsection}%
    {2}% level
    {0em}% indent
    {0.75cm \@plus 0.1ex \@minus -0.05ex}% Beforeskip
    {0.25cm \@plus 0.2em}% Afterskip
    {\bf\large}% Style
}

\renewcommand{\subsubsection}{%
  \@startsection{subsubsection}%
    {3}% level
    {0em}% indent
    {0.375cm \@plus 0.1ex \@minus -0.05ex}% Beforeskip
    {-0.25cm \@plus 0.2em}% Afterskip
    {\normalfont\normalsize\bfseries}% Style
}

\renewcommand{\paragraph}[1]{%
  \par% ensure starting on a new paragraph
  \addvspace{\medskipamount}% some vertical space
  \noindent\textit{#1\@addpunct{.}}\quad\ignorespaces
}

\makeatother

\theoremstyle{definition}

\theoremstyle{plain}
\newtheorem{note}{Remark}
\newtheorem{theo}{Theorem}[section]

\newtheorem{prop}[theo]{Proposition}

\numberwithin{equation}{section}
\numberwithin{defn}{section}
\numberwithin{note}{section}

\stackMath

\let\oldker\ker
\let\oldforall\forall
\let\oldexists\exists
\renewcommand\widehat[1]{%
\savestack{\tmpbox}{\stretchto{%
    \scaleto{%
        \scalerel*[\widthof{\ensuremath{#1}}]{\kern.1pt\mathchar"0362\kern.1pt}%
        {\rule{0ex}{\textheight}}%WIDTH-LIMITED CIRCUMFLEX
    }{\textheight}% 
}{2.4ex}}%
\stackon[-6.9pt]{#1}{\tmpbox}%
}
\providecommand{\hide}[1]{}

\setstcolor{red}

\ProvideDocumentCommand \comment {o m m}{%
    \hl{#2}\footnote{%
        \IfValueT{#1}{#1:} \textcolor{red}{#3.}%
    }%
}

\RenewDocumentCommand \newline {o}{%
\hfill\\[\IfValueTF{#1}{#1}{0} pt]
}
\renewcommand{\-}{\mspace{-1.5mu}}
\newcommand{\+}{\mspace{1.5mu}}

\newcommand{\n}{\noindent}
\newcommand{\vs}{\vspace{0.5cm}}

\newcommand{\f}{\frac}
\newcommand{\be}{\begin{equation}}
\newcommand{\ee}{\end{equation}}
\newcommand{\ie}{\emph{i.e.}}
\newcommand{\eg}{\emph{e.g.}}

\newcommand \inlineFrac[2]{%
    {{}^{#1}}\!/\-{{}_#2}%
}
\newcommand{\pInfty}{ {\scriptstyle +}\+\infty}

\renewcommand{\forall}{\oldforall\,}
\RenewDocumentCommand \exists {s}{%
    \IfBooleanTF{#1}{\oldexists!}{\oldexists} \;%
}

\RenewDocumentCommand \to {o}{%
    \IfValueTF{#1}{%
        \xrightarrow[\,#1\,]{\;}%
    }{%
        \,\rightarrow\,%
    }%
}

\ProvideDocumentCommand \conjugate {s m}{%
    \IfBooleanTF{#1}{%
        \overline{#2}%
    }{%
        \bar{#2}%
    }%
}
\NewDocumentCommand \abs {s m}{%
    \IfBooleanTF{#1}{\left\lvert#2\right\rvert}{\lvert#2\rvert}%
}
\renewcommand*{\vec}[1]{\boldsymbol{#1}}

\NewDocumentCommand \hilbert{s}{%
    \IfBooleanTF{#1}{\mathcal{H}}{\mathscr{H}}%
}
\NewDocumentCommand \X{s}{%
    \IfBooleanTF{#1}{\mathcal{X}}{\mathfrak{X}}%
}
\NewDocumentCommand \scalar { m m o }{%
    \langle#1,\,#2\rangle\IfValueT{#3}{_{#3}}%
}
\NewDocumentCommand \norm { s m o }{%
    \IfBooleanTF{#1}{\left\lVert#2\right\rVert}{\lVert#2\rVert}\IfValueT{#3}{_{#3}}%
}

\renewcommand*{\ker}[1]{\oldker(#1)}
\NewDocumentCommand \tr {s o}{%
    \mathop{\mathrm{Tr}}\IfValueT{#2}{%
        \IfBooleanTF{#1}{%
            \-\left(#2\right)%
        }{%
        \+(#2)%
        }%
    }%
}

\NewDocumentCommand \resolvent {m o}{%
    \mathcal{R}_{#1}\IfValueT{#2}{(#2)}%
}

\NewDocumentCommand \FT { s m o }{
    \IfBooleanTF{#1}{ \IfValueTF{#3}{ (\mathcal{F} \+  #2)\-(#3) }{%
    \mathcal{F} \+  #2 } }{ \hat{#2}\IfValueT{#3}{(#3)} }
}

\newcommand{\R}{\mathbb{R}}
\newcommand{\Rplus}{\mathbb{R}_{+}}
\newcommand{\N}{\mathbb{N}}

\NewDocumentCommand \Z {o}{%
    \IfValueTF{#1}{%
        \mathbb{Z}_{\,#1}%
    }{%
        \mathbb{Z}%
    }%
}

\allowdisplaybreaks

\title[Efimov Effect in the Born-Oppenheimer Approximation]{A Zero-Range Model for the Efimov Effect\\ in the Born-Oppenheimer Approximation}

\author[G. Basti]{Giulia Basti}
\address{\vspace{-0.25cm}\small Università degli studi di Roma \emph{La Sapienza}, Piazzale Aldo Moro 5 - 00185, Rome (IT)}
\author[D. Ferretti]{Daniele Ferretti}
\address{\vspace{-0.25cm}\small \emph{Gran Sasso Science Institute}, Via Michele Iacobucci 2 - 67100, L'Aquila (IT)}
\author[A. Teta]{Alessandro Teta}
\address{\vspace{-0.25cm}\small Università degli studi di Roma \emph{La Sapienza}, Piazzale Aldo Moro 5 - 00185, Rome (IT)\\{\footnotesize \textsf{teta@math.uniroma1.it} }
}

\thanks{The authors acknowledge the support of the GNFM Gruppo Nazionale per la Fisica Matematica - INdAM.\newline
D. Ferretti is financially supported by the European Research Council through the ERC Stg MaTCh, grant agreement no. 101117299.}

\begin{document}

\begin{abstract}
In this note we discuss the Efimov effect emerging in a  three-particle quantum system with zero-range interactions. In particular, we consider two non-interacting identical bosons plus a different lighter particle such that the interaction between a boson and the light particle is resonant. We also assume the validity of the Born-Oppenheimer  approximation. Under these conditions we show that the three-particle system exhibits infinitely many negative eigenvalues which accumulate at zero and satisfy the universal geometrical law characterising the Efimov effect. 

\n
The result we find is a generalisation of  previous results   recently obtained in \cite{FST, S}.
\newline[10]
\begin{footnotesize}

\end{footnotesize}
\end{abstract}

\maketitle

%-------------------------------------------
%  ######  ########  ######            ##    
% ##    ## ##       ##    ##         ####   
% ##       ##       ##                 ##   
%  ######  ######   ##                 ##   
%       ## ##       ##                 ##   
% ##    ## ##       ##    ## ###       ##   
%  ######  ########  ######  ###     ###### 
%-------------------------------------------

\section{Introduction}

\n
The Efimov effect is an interesting physical phenomenon occurring in three-particle quantum systems in three dimensions~\cite{efimov1,efimov2}.
It is characterised by the appearance of an infinite sequence of negative eigenvalues $E_n$ of the three-body Hamiltonian, where $E_n \to 0$ as $ n \to \infty$, under the specific condition that the two-particle subsystems have no bound states and at least two of them exhibit a zero-energy resonance or, equivalently, an infinite two-body scattering length.
For a review, see \eg~\cite{NE}.\newline
%It should be stressed that the effect is independent of the details of the two-body interactions and it is 
A remarkable feature of this effect is that the distribution of the eigenvalues satisfies the universal geometrical law
\begin{equation}\label{gela}
    \frac{E_{n}}{E_{n+1}} \:\to\: e^{\frac{2\pi}{s}} \qquad \mbox{for\; $n \to \infty$}\,,
\end{equation}
where the parameter $s>0$ depends only on the mass ratios and possibly on the statistics of the particles.

%\vs
%
%\n
%According to an intuitive physical picture, the  three-particle bound states (or trimers) associated to the eigenvalues are determined by a long range, attractive effective interaction of kinetic origin, which is produced by the resonance condition and does not depend on the details of the two-body potentials.
%Roughly speaking, in a trimer the attraction between two particles is mediated by the third one, which is moving back and forth between the two.
%It should also be stressed that the Efimov effect disappears if the two-body potentials become more attractive causing the destruction of the zero-energy resonance.  
%
%\vs

\n
For the first experimental evidence of Efimov quantum states we  refer to~\cite{kra}.

%\n
%It is worth recalling that just one year after the publication of Efimov's seminal works, Faddeev suggested an argument for the derivation of~\eqref{gela} based on the direct inspection of the three-body resolvent operator in the low-energy regime (see~\cite{Fa71}, \cite[\S 3.4.2]{FaMe93} and~\cite{ALM98}).

\n
The mathematical analysis of the Efimov effect was initiated by Yafaev in 1974~\cite{yafaev}.
He studied a symmetrised form of the Faddeev equations for the bound states of the three-particle Hamiltonian and proved the existence of an infinite number of negative eigenvalues.
In 1993  Sobolev~\cite{sobolev} used a slightly different symmetrization of the equations and proved the following asymptotic behaviour
\begin{equation}\label{asob}
    \lim_{z \to 0^-} \frac{N\-(z)}{ \abs{\log\abs{z}\mspace{0.75mu}}} = \frac{s}{2\pi}  \,,
\end{equation}
where $N\-(z)$ denotes the number of eigenvalues smaller than $z<0$.
Note that~\eqref{asob} is consistent with the law~\eqref{gela}.
In the same year  Tamura~\cite{tamura2} obtained the same result under more general conditions on the two-body potentials.
Other mathematical proofs of the effect were obtained by Ovchinnikov and Sigal in 1979~\cite{OS} and Tamura in 1991~\cite{tamura1} using a variational approach based on the Born-Oppenheimer approximation. 
%Let us also mention that an infinite sequence of eigenvalues fulfilling a relation of the form~\eqref{asob} was found by Lakaev~\cite{La93} for a system of three quantum particles moving on the lattice $\mathbb{Z}^3$ (see also~\cite{ALM04}). 
Further results on the subject have been obtained in \cite{gridnev2}, in~\cite{BT}, for the case of two identical fermions and a different particle, and in~\cite{gridnev1}, for a two-dimensional variant of the problem.
We stress that a rigorous derivation of the law~\eqref{gela} is lacking in the mathematical works mentioned above.

\n
Regarding the Efimov effect, it is also worth mentioning the work of % that, before the seminal works of Efimov, 
Minlos and Faddeev (\cite{MF1, MF2}) on the problem of defining the Hamiltonian for a system of three bosons with zero-range interactions in three dimensions. %It was known that such Hamiltonian cannot be defined considering only pairwise zero-range interactions. 
They constructed a self-adjoint  Hamiltonian by imposing suitable two-body boundary conditions on the coincidence hyperplanes, {\it i.e.}, where any two particle positions coincide,  and a three-body boundary condition at the triple-coincidence point, {\it i.e.}, where all three particle positions coincide.
They also proved that such a Hamiltonian is unbounded from below due to the presence of an infinite sequence of negative eigenvalues diverging to $-\infty$.
%Such instability property can be seen as a fall to the center phenomenon and it is due to the fact that the interaction becomes too strong and attractive when the three particles are very close to each other.
As a further interesting result, in the case of infinite two-body scattering length (corresponding to the resonant case), they proved the occurrence of the  Efimov effect with a rigorous derivation of the law~\eqref{gela} (note that the works of Minlos and Faddeev predate those of Efimov).
%A further interesting result of the analysis of Minlos and Faddeev, even if it is not explicitly emphasized, is the proof of the Efimov effect in the case of infinite two-body scattering length (corresponding to the resonant case), with a rigorous derivation of the law~\eqref{gela}. This in particular shows that the occurrence of the Efimov effect can be obtained also with zero-range interactions, the only crucial condition being the presence of an infinite two-body scattering length.
Nevertheless, such a result is somewhat tainted by the fact that the Hamiltonian is unbounded from below and therefore unsatisfactory from the physical point of view.

%\n
%Our aim is to present a mathematical proof of the Efimov effect and law~\eqref{gela} for a bounded from below Hamiltonian obtained by a slight modification of the Minlos and Faddeev Hamiltonian.
%
%\vs

\n
We mention that the problem of constructing a lower bounded Hamiltonian for a three-body system with zero-range interactions has been recently approached in the literature (see, \eg,~\cite{BCFT, FiTe, miche} and, for the many particle case,~\cite{FT1, FT2}).
The idea is to introduce an effective three-body force acting only when the three particles are close to each other, preventing the fall-to-the-centre phenomenon.

\n
In~\cite{FFT}, the role of the effective three-body force is played by a three-body hard-core repulsion.
It is shown that the Hamiltonian is self-adjoint and bounded from below; moreover, it is also proved that  the Efimov effect occurs in the resonant case, \ie, there exists an infinite sequence of negative eigenvalues satisfying~\eqref{gela}.

\n
In this note, we discuss the approach to the Efimov effect in a three-particle system with zero-range interactions by using the Born-Oppenheimer approximation (for recent results in this direction, see~\cite{FST, FiTe, S}; see also~\cite{FRS} for a result in a model with separable potentials, and~\cite{CPS} for the analysis of the Born-Oppenheimer approximation in the one dimensional case).  

\n
More precisely, we consider a system made of two identical scalar bosons with mass $M$ and another different spinless particle with mass $m$ %, with $m \ll M$ 
%\n
%Our work can be viewed as an attempt to make rigorous the original physical argument of Efimov. Indeed, Efimov takes into account three identical bosons and his approach is based on the replacement of the two-body potential with a boundary condition, which is essentially equivalent to consider a two-body zero-range interaction. Then, he introduces hyper-spherical coordinates and shows that if the two-body scattering length is infinite then the problem becomes separable and in the equation for the hyper-radius $R$ the long range, attractive effective  potential $- (s_0^2 +1/4)/R^2$ appears. The behavior for small $R$ of this potential is too singular and an extra boundary condition at short distance  must be imposed. After this ad hoc procedure, he obtains the infinite sequence of negative eigenvalues satisfying the law~\eqref{gela} as a consequence of the large $R$ behavior of the effective potential.
%
%\n
%The self-adjoint and bounded from below Hamiltonian constructed in this paper can be considered as the rigorous counterpart of the ad hoc regularization scheme mentioned above. Furthermore, we show that the eigenvalues and eigenvectors found in a formal way in the physical literature are in fact eigenvalues and eigenvectors of our Hamiltonian in a rigorous sense and, accordingly, we obtain a mathematical proof of~\eqref{gela}. 
and let $\vec{x}_1, \vec{x}_2, \vec{x}_3 \in \mathbb{R}^3$ be the coordinates of the bosons and the other particle, respectively.
We introduce the system of Jacobi coordinates $\vec{r}_{\!\mathrm{cm}},\vec{x},\vec{y} \in \mathbb{R}^3$ defined as
\begin{equation*}
		\vec{r}_{\!\mathrm{cm}} \vcentcolon= \frac{M (\vec{x}_1 \- + \vec{x}_2) + m\, \vec{x}_3}{2M\- +m}\,, \qquad 
		\vec{y} \vcentcolon= \vec{x}_1\- - \vec{x}_2\,, \qquad 
		\vec{x} \vcentcolon=  \vec{x}_3 - \frac{\vec{x}_1\- + \vec{x}_2}{2}\, .
	\end{equation*}

\n
We study the problem in  the centre-of-mass reference frame and we set $\hbar=1$.
The Hilbert space encoding the bosonic symmetry of the system is the following
\begin{equation}\label{L2}
    L^2_{s}(\R^6) := \left\{ \psi \-\in\- L^2(\R^6)  \:\big\vert\;  \psi(\vec{x},\vec{y}) = \psi(\vec{x},-\vec{y}) \- \right\}. 
\end{equation}
For simplicity, we also assume that the two identical bosons do not interact.
Then the heuristic Hamiltonian describing our three-particle system is expressed by
\begin{equation}\label{eq:Hxy}
    \widetilde{H} =  - \frac{1}{\mu}\Delta_{\vec{y}} -\frac{1}{\nu} \Delta_{\vec{x}} +   \delta (\vec{x}- \vec{y}/2) +  \delta (\vec{x} + \vec{y}/2),
\end{equation}

\n
where
\be
    \mu = M, \qquad \nu=\f{4Mm}{2M\-+m}\,.
\ee

\n
The first step is to give a rigorous meaning to the above formal Hamiltonian on the Hilbert space
$L^2_{s}(\R^6)$.
By definition, the Hamiltonian is a self-adjoint extension of the symmetric operator
\begin{align}
    \dot{H}_0 =  - \frac{1}{\mu}\Delta_{\vec{x}} -\frac{1}{\nu} \Delta_{\vec{y}}\,, && D(\dot{H}_0) = \left\{ \psi\- \in\- L^2_{s}(\R^6) \cap H^2(\R^6) \:\big\vert\;  \psi|_{\pi_{\pm}}\! =0 \right\},
\end{align}
where
\begin{equation}
    \pi_{\pm} \vcentcolon= \left\{\- (\vec{x}, \vec{y} )\- \in \R^6 \;\big\vert\; \vec{x}= \pm \+ \vec{y} /2 \right\}\-.
\end{equation}
Using the results in~\cite{FT1}, we consider the bounded from below Hamiltonian $H, D(H)$ in $L^2_{s}(\R^6)$ satisfying the two conditions:

\begin{itemize}
    \item[(C-1)] $\;H \psi = H_0 \psi$, where $H_0$ is the free three-body Hamiltonian,  for  $\psi \in D(H)$ with $\psi|_{\pi_{\pm}}\!=0$;
    \item[(C-2)] $\;\psi \in D(H)$ satisfies the boundary condition on $\pi_{\pm}$

    \begin{equation}\label{BC}
        \psi(\vec{x},\vec{y})= \f{\xi(\pm \vec{y})}{\abs{\vec{x} \mp \vec{y}/2} } + \Big( \alpha + \f{\theta(\abs{\vec{y}})}{\abs{\vec{y}}} \Big) \,\xi(\pm \vec{y} ) + o(1) \;\;\;\;\;\;\;\; \text{for}\;\; \vec{x} \to \pm \vec{y}/2
    \end{equation}
    \n
    where
    \begin{equation}
        \xi(\vec{y}) \-\vcentcolon=\! \lim\limits_{\vec{x} \to \vec{y}/2}\; \abs{\vec{x}\-- \vec{y} /2 }\, \psi(\vec{x}, \vec{y}),
    \end{equation}
    \n $\alpha\-\in\-\R$, and $\theta \,:\, \R^+ \longrightarrow \R$ is smooth and satisfies the conditions
    $$\theta(0)=1\, ,  \;\;\;\;\;\; \theta(r)=0 \;\;\; \text{ for} \;\,\; r\geq r_0>0.$$
\end{itemize}
The parameter $\alpha$ is the inverse of the two-body scattering length.
Moreover, the role of the function $\theta/\abs{\+\cdot\+}$ is to regularise the interaction when the positions of the three particles coincide, thereby preventing the fall-to-centre phenomenon (for additional technical details, we refer to~\cite{FT1}).
Notice that $r_0$ can be chosen to be arbitrarily small.

\n
Our aim is to study the eigenvalue problem 
\begin{equation}
    H \Psi = E \+\Psi \,, \qquad E<0
\end{equation}
and the occurrence of the Efimov effect for the Hamiltonian $H, D(H)$ under the two assumptions:

\begin{itemize}
    \item[-] $\alpha=0$, \ie, for an infinite two-body scattering length,
    \item[-] $M\gg m$ (equivalently, $\mu\gg \nu$).
\end{itemize}

\n
More precisely, in this note we assume the validity of the Born-Oppenheimer approximation, which means that 

\begin{equation}
    E \simeq E_{\mathrm{BO}} \,, \qquad \Psi (\vec{x},\vec{y}) \simeq \zeta (\vec{y})\+  \phi( \vec{x}; \vec{y} )
\end{equation}

\n
where, for any  fixed $\vec{y}$, $\phi(\vec{\cdot}\,; \vec{y}) $ is an eigenvector with eigenvalue $\mathcal E (\abs{\vec{y}})$ of the formal one-body Hamiltonian with two point interactions placed at $\pm \vec{y}/2$
\begin{equation}\label{hfast}
  \widetilde{h}(\vec{y})= -\frac{1}{\nu} \Delta + \delta (\+\vec{\cdot}-\vec{y}/2) +  \delta (\+\vec{\cdot} + \vec{y}/2),
\end{equation}
and $\zeta$ is an eigenvector with eigenvalue $E_{\mathrm{BO}}$ of the one-body Hamiltonian 

\be\label{hslow}
    h_{\mathcal E} = -\f{1}{\mu} \Delta + \mathcal E(\abs{\+\vec{\cdot}\+}).
\ee

\n
Hence, the problem is reduced to first solving the eigenvalue problem for the fast dynamics associated with~\eqref{hfast}, and then treating the resulting eigenvalue $\mathcal E(\abs{\vec{y}})$ as an effective potential for the slow dynamics described by~\eqref{hslow}.

\n
It is important to stress that the one-body Hamiltonian $\widetilde{h}(\vec{y})$ is obtained by freezing the slow dynamics in the three-body Hamiltonian $H$ (for $\alpha=0$) and then  considering the coordinate $\vec{y}$ as a fixed parameter.
Hence, the rigorous counterpart $h(\vec{y}), D( h(\vec{y}))$ of~\eqref{hfast} is characterised as the self-adjoint and bounded from below operator in $L^2(\R^3)$ satisfying the two conditions:

\begin{itemize}
    \item[(c-1)] $\; h(\vec{y}) u = h_0  u$, where $h_0=-\frac{1}{\nu}\Delta$ is the free one-body Hamiltonian, for $u \in D(h(\vec{y}))$ with $u (\pm \vec{y}/2)=0$; 
    \item[(c-2)]  $\;u \in D(h(\vec{y}))$ satisfies the boundary condition at $\vec{x}= \pm \vec{y}/2$
    \begin{equation}\label{bc}
        u(\vec{x})= \f{q}{\abs{\vec{x} \mp \vec{y}/2} } +  \f{\theta(\abs{\vec{y}})}{\abs{\vec{y}}}  q + o(1) \;\;\;\;\;\;\;\; \text{for}\;\;  \vec{x} \to \pm \vec{y}/2
    \end{equation}
    where
    \begin{equation}\label{bc1}
    q\-\vcentcolon=\! \lim\limits_{\vec{x} \to \vec{y}/2}\; \abs{\vec{x}\--  \vec{y}/2 }\, u(\vec{x}).
    \end{equation}
\end{itemize}

\n
Such an operator is known as a Hamiltonian with non local point interactions at the points $\pm \vec{y}/2$, and its main properties are recalled in the next section.

\n
The eigenvalues $E_{\mathrm{BO}}$ of  the one-body Hamiltonian $h_{\mathcal E}$ are usually assumed to be a good approximation  of the true eigenvalues $E$ of the three-body Hamiltonian $H, D(H)$ for $M \gg m$.
A rigorous proof of this fact is not discussed in this note, and it will be addressed in a future work.

\n
The next theorem summarises our main result. 

\begin{theo}\label{th1} The following two statements hold:

\n
\begin{enumerate}[label=(\roman*)]
    \item for any fixed $\vec{y} \in \R^3$, the Hamiltonian $h(\vec{y})$ admits the negative eigenvalue
    \begin{equation}\label{aule}
        \mathcal E(|\vec{y}|)=-\f{ \Big(W\-\big(e^{\theta(|\vec{y}|)}) - \theta(|\vec{y}| \big) \Big)^{\!\!2} }{{\nu |\vec{y}|^2}}
    \end{equation}
    \n where $W: \big[\!-\-\frac{1}{e},\infty\big)\longrightarrow [-1,\infty)$ denotes the principal branch of the Lambert function; %the corresponding eigenfunction is 
%\be
%\phi(\vec{x} ; \vec{y}) = C\, \Big(\,\f{e^{- \sqrt{ \nu |\mathcal E ( |\vec{y}|) | } \,|\vec{x}-\vec{y}/2|} }{|\vec{x}- \vec{y}/2| } + \f{e^{- \sqrt{ \nu |\mathcal E ( |\vec{y}|) | } \,|\vec{x}+\vec{y}/2|} }{|\vec{x}+ \vec{y}/2| } \,\Big)
%\ee
%
%\n
%where $C$ is a normalisation constant.

\item the Hamiltonian $h_{\mathcal E}$ admits an infinite sequence of negative eigenvalues $\{E_{\mathrm{BO};\,n}\}$, $n \in \N$, such that  $\lim\limits_{n\to\infty} E_{\mathrm{BO};\,n}=0$ and 

\begin{equation}\label{lege}
    \lim_{n \to \infty} \f{E_{\mathrm{BO};\,n} }{E_{\mathrm{BO};\,n+1}}= \, e^{\f{2\pi}{\beta}},
\end{equation}
\n
where $\, \beta= \sqrt{\f{\mu}{\nu} W(1)^2 - \f{1}{4} }\; $ and $\;W(1) \simeq 0.5671$.
\end{enumerate}
\end{theo}

\begin{note}
The result expressed in Theorem \ref{th1} is essentially the same found in \cite{FST, S}. The only difference is that in these latter works our function $\theta$ is replaced by the function
$$g(r)=e^{-r/\sqrt{2}} \Big( \sin \f{r}{\sqrt{2}} + \cos \f{r}{\sqrt{2}} \Big)\,. 
$$
Such a function emerges a specific choice in the construction of the one-body Hamiltonian $h(\vec{y})$ with (non-local) point interactions placed at $\vec{x}=\pm\vec{y}/2$ in the context of the theory of self-adjoint extensions.
On the other hand, in this note the Hamiltonian $h(\vec{y})$ is obtained as the limit for $M\to \infty$ of the three-body Hamiltonian $H$.
In this sense we believe that our approach here is more natural from the physical point of view and, moreover, the result is slightly  more general, since it holds for any choice of the function $\theta$ satisfying the properties stated above.
\end{note}

\n
The proof of Theorem \ref{th1} will be given in the following two sections. 

\n
In Section~\ref{sec:fast} we recall definition and main properties of the Hamiltonian $h(\vec{y})$ and then solve the corresponding eigenvalue problem. In particular, we compute the eigenvalue $\mathcal E(|\vec{y}|)$ and the corresponding eigenfunction.

\n
In Section~\ref{sec:slow} we study the eigenvalue problem for $h_{\mathcal E}$, showing that there is an infinite sequence of negative eigenvalues accumulating at zero and satisfying the asymptotic law~\eqref{lege}. The corresponding eigenfunctions are also explicitly characterised.

%---------------------------------------------
%  ######  ########  ######          #######  
% ##    ## ##       ##    ##        ##     ## 
% ##       ##       ##                     ## 
%  ######  ######   ##               #######  
%       ## ##       ##              ##        
% ##    ## ##       ##    ## ###    ##        
%  ######  ########  ######  ###    ######### 
%---------------------------------------------

\section{Fast Dynamics}\label{sec:fast}

\n
We start this section giving the rigorous definition of the operator formally written in \eqref{hfast}. 
Let 

$$G^\lambda(\vec{x})=\frac{e^{-\sqrt{\lambda\nu}\+|\vec{x}|}}{|\vec{x}|}$$

\n
 for $\lambda>0$, be the solution to $\;(-\frac{1}{\nu}\Delta+\lambda)G^\lambda=\frac{4\pi}{\nu}\delta_{\vec{0}} \;$ normalised so that it has the desired singular behaviour
 
\[
	G^\lambda(\vec{x})=\frac1{|\vec{x}|}-\sqrt{\lambda\nu}+o(1)
\]

\n
as $\vec{x}\to 0.$ Then (see \cite{albeverio}, \cite{DG}, \cite{FST}), the operator $h(\vec{y})$ has domain

\begin{equation}\label{eq:dom_fast}
	\begin{aligned}
		D(h(\vec{y}))=\bigg\{&u\in L^2(\mathbb{R}^3) : u=\omega^\lambda+q \, G^\lambda(\+\vec{\cdot}+\vec{y}/2)+q\, G^\lambda(\+\vec{\cdot}-\vec{y}/2), \;  q\in \mathbb{C},\; \omega^\lambda\in H^2(\mathbb{R}^3), \\
		&\omega^\lambda\big(\pm \vec{y}/2\big)=q\Big(\frac{\theta(|\vec{y}|)}{|\vec{y}|}+\sqrt{\lambda\nu}-G^\lambda( \vec{y})\Big)\bigg\}\,.
	\end{aligned}
\end{equation}

\n
The action of $h(\vec{y})$ on $u\in D(h(\vec{y}))$ is given by

\begin{equation}\label{eq:act_fast}
	(h(\vec{y})+\lambda)u=(h_0+\lambda)\omega^\lambda
\end{equation}

\n
where $h_0=-\frac1\nu\Delta$. Notice that 
the last equality in \eqref{eq:dom_fast} rephrases the boundary condition \eqref{bc}, \eqref{bc1} and, moreover, if $u \in D(h(\vec{y}))$ and $u(\pm \vec{y}/2)=0$ then  \eqref{eq:act_fast} implies  $h(\vec{y}) u=h_0 u$. Hence the  two conditions (c-1), (c-2) stated in the Introduction are satisfied.

\n
The operator in $L^2(\R^3)$ defined by \eqref{eq:dom_fast}, \eqref{eq:act_fast} is self-adjoint and bounded from below and it is known in the literature as a Schr\"odinger operator with non local point interactions at $\pm \vec{y}/2$. The non local character is due to the presence of the function $\theta$ which makes  the boundary condition, and therefore the strength of the interaction, dependent on the distance of the two point interactions. Notice that the form chosen for the function $\theta$ is such that the boundary condition remains well defined in the limit $|\vec{y}| \to 0$ (whereas if $\theta=0$ it loses its meaning in the limit). 

\n
The Hamiltonian $h(\vec{y})$ defines a solvable model, in the sense that the spectrum and the (proper and generalised) eigenfunctions can be explicitly characterised. 
In particular, it is easy to check that $h(\vec{y})$ does not have any positive eigenvalues. Let's now show the existence of a negative eigenvalue. Assume $-\lambda<0$ is an eigenvalue and let $\phi=\omega^\lambda+q \Big(G^\lambda(\+\vec{\cdot}+\vec{y}/2)+G^\lambda(\+\vec{\cdot}-\vec{y}/2)\Big)\in D(h(\vec{y}))$ be the corresponding eigenvector. Equation \eqref{eq:act_fast} immediately yields $(h_0+\lambda)\omega^\lambda=0$, and therefore $\omega^\lambda = 0$.
Thus, by the boundary condition in \eqref{eq:dom_fast} we conclude that $\lambda$ is a solution of the equation 

\[
	\frac{\theta(|\vec{y}|)}{|\vec{y}|}+\sqrt{\nu\lambda}-G^\lambda(\vec{y})=0.
\]

\n
Introducing the new variable $s=\sqrt{\nu\lambda}|\vec{y}|+\theta(|\vec{y}|)$ and recalling the expression of $G^\lambda$ we rewrite the equation as

\[
	se^s=e^{\theta(|\vec{y}|)}\,.
\]

\n
Since $e^{\theta(|\vec{y}|)}>0$, the only solution to the above equation is given by $s=W(e^{\theta(|\vec{y}|)})$ with $W$ the principal branch of the Lambert function defined as $W=f^{-1}$ with $f(s) = s\+e^s, s\geq -1$.
Note that $W$ is smooth, non-negative, and increasing, with $W(e)=1$.
We conclude that the only eigenvalue of the operator $h(\vec{y})$ is

 \[
 	\mathcal{E}(|\vec{y}|)=-\lambda=-\frac{\Big( W(e^{\theta(\abs{\vec{y}})})-\theta(\abs{\vec{y}}) \Big)^2}{\nu|\vec{y}|^2}.
\]

\n
  In particular, recalling the assumption on $\theta$, we have that $\; \mathcal{E}(|\vec{y}|)=-\frac{W(1)^2}{\nu |\vec{y}|^2}\; $ for $\; |\vec{y}|\geq r_0.$ Moreover, for $\; |\vec{y}|<r_0\; $ we have that $\;\mathcal{E}(|\vec{y}|),\,$ whose exact expression depends on $\;\theta,$ is always continuous and bounded.
  The corresponding eigenvector is
  
  \[
  	\phi(\+\vec{\cdot}\+; \vec{y})=q\Big(G^{-\mathcal{E}(\vec{|y|})}(\+\vec{\cdot}+{\vec{y}}/{2})+G^{-\mathcal{E}(\vec{|y|})}(\+\vec{\cdot}-{\vec{y}}/{2})\Big)
  \]
  
  \n
  with $q$ normalising constant. 
  
\n
It is important to stress once again the regularising role of the function $\theta$. Indeed, in the case $\theta=0$  the eigenvalue would reduce to  $\; \mathcal{E}(|\vec{y}|)=-\frac{W(1)^2}{\nu |\vec{y}|^2}\; $ for any $\; \vec{y} \in \R^3$ and therefore the potential $\;\mathcal E(\abs{\vec{y}})\,$ for the slow dynamics would be too singular for $\vec{y} \to 0$.

\section{Slow Dynamics}\label{sec:slow}

\n
Here we study the eigenvalue problem for the one-body Hamiltonian 

$$
h_{\mathcal E}= -\f{1}{\mu} \Delta + \mathcal E (|\cdot|)\,, \;\;\;\;\;\; D(h_{\mathcal E}) = H^2(\R^3)
$$

\n
where $\mathcal E (|\cdot|)$ is the continuous and bounded function given by \eqref{aule}. We restrict the analysis to the $s$-wave sector and therefore the problem to be solved is 

%Given $r_0\in\Rplus$ such that $\theta(r)=0$ for all $r>r_0$, and $\mu>\frac{\nu}{4\+W(1)^2}$, let $v:[0,r_0]\longrightarrow \R$ be a bounded, continuous function satisfying $v(r_0)=-\frac{\mu \+W(1)^2\!\-}{\nu}\frac{1}{r_0^2}$ defined by
%\begin{equation}
%    \mu\,\mathcal{E}(\vec{y})=\begin{dcases}
%        v(\abs{\vec{y}}),\qquad & \abs{\vec{y}}\in [0, r_0],\\
%        -\frac{\mu \+W(1)^2\!\-}{\nu}\,\frac{1}{\abs{\vec{y}}^2},\qquad &\abs{\vec{y}} > r_0.
%    \end{dcases}
%\end{equation}
%We are interested in the eigenvalue problem for the Schr{\"o}dinger operator $-\frac{1}{\mu}\Delta + \mathcal{E}, H^2(\R^3)$ in the $s$-wave sector:

\begin{equation}\label{eq:sWaveEigenvalueProblem}
    \begin{dcases}
        -\zeta''(r)-\frac{2}{r}\+\zeta'(r)+\mu\, \mathcal{E}(r)\+\zeta(r)=\mu\+ E_{\mathrm{BO}}\,\zeta(r),\qquad \min \mathcal{E}\leq E_{\mathrm{BO}}< 0, \, r\in\Rplus,\\
        r\,\zeta(r)\in H^2(\Rplus).
    \end{dcases}
\end{equation}

\n
Recall that $H^2(\R^3)\subset C^{\+0}(\R^3)$: specifically, $\zeta$ is finite at every point.\newline
Consider the substitution $u\-:\, r\:\longmapsto r\, \zeta (r) \!\in\! H^2(\Rplus)\cap H^1_0(\Rplus)$ and the notation $\lambda\vcentcolon=\sqrt{-\mu\+ E_{\mathrm{BO}}}$, and $\beta\vcentcolon=\-\sqrt{\frac{\mu \+W(1)^2\!\-}{\nu}-\frac{1}{4}\+}\+$.
Then,~\eqref{eq:sWaveEigenvalueProblem} reads
\begin{equation}\label{eq:entireProblem}
    \begin{dcases}
        u''-(\mu \,\mathcal{E}+\lambda^2)\+u=0,\\
        u\in H^2(\Rplus)\cap H^1_0(\Rplus).
    \end{dcases}
\end{equation}

\n
In other words, denoting 

$$v(r)= \mu\, \mathcal E (r)\, ,\;\;\;\;\;\;\;\; 0 \leq r \leq  r_0
$$

\n
we arrive at 

%\begin{subequations}
\begin{align}
 &   u''(r)-\big(v(r)+\lambda^2\big)\+u(r)=0,  & 0< r \leq r_0,\label{eq:innerProblem}\\
   & u''(r)+\frac{\beta^2\-+\inlineFrac{1}{4}}{r^2} \,u(r)-\lambda^2\+ u(r)=0, & r>r_0,\label{eq:exteriorProblem}
\end{align}
%\end{subequations}

\n
with the boundary condition $u(0)=0$.\newline
Because of the continuity of $v$, one can find that $\lambda\longmapsto w_\lambda\in C^{\+0}\mspace{-0.75mu}\big(\Rplus,\mspace{0.75mu} C^2([0,r_0])\mspace{-0.75mu}\big)$ is a solution to equation~\eqref{eq:innerProblem} with $w_\lambda(0)=0$ and\footnote{Observe that the value of $w_\lambda'$ at zero corresponds to that of the function $\zeta$ at the origin.} $w_\lambda'(0)=c\neq 0$.
Such a solution exists and is unique by the Picard–Lindel\"of theorem.
In particular, it follows that
$$w_\lambda(r)=w_{\+0}(r)+R_\lambda(r),\qquad \text{with }\,\lim_{\lambda\to 0^+}\norm{R_\lambda}[C^2([0,r_0])]=0.$$
As an explicit example, consider $v\equiv -\lambda^2_0\vcentcolon=-\frac{\beta^2+\,\inlineFrac{1}{4}}{r_0^2}$; in this case, one has $0<\lambda\leq \lambda_0$ and
$$w_\lambda(r)=\begin{dcases}
    c\, r,\quad & \lambda=\lambda_0,\\
    \frac{c\,\sin(\lambda_0 \+ r)}{\lambda_0}+R_\lambda(r),\quad & \lambda< \lambda_0,
\end{dcases}\qquad \text{with }\, R_\lambda(r)=\frac{c\,\sin\!\big(\-\sqrt{\lambda_0^2-\-\lambda^2\+}\+ r\big)}{\sqrt{\lambda_0^2-\-\lambda^2\+}}-\frac{c\,\sin(\lambda_0 \+ r)}{\lambda_0}.$$
Next, the second-order ODE~\eqref{eq:exteriorProblem} can be solved explicitly; moreover, only one of the independent solutions decays at infinity, \ie
$$r\longmapsto \sqrt{r\+}\+K_{i\beta}(\lambda\+ r).$$
Here, $K_{i\beta}$ is the modified Bessel function of the second kind, sometimes also known as the Macdonald function, which satisfies the following asymptotic behaviours (see~\cite[eq. 10.40.2 p. 255, eq. 10.45.7 p. 261]{OlMa10})
\begin{subequations}
\begin{gather}
    K_{i\beta}(z)=\sqrt{\-\tfrac{\pi}{2\+z}}\,e^{-z}\!\left[1-\tfrac{4\beta^2+1}{8\+z}+O(z^{-2})\right]\!,\qquad \text{for }\,z\to \pInfty,\label{eq:infinityMacdonald}\\
    K_{i \beta}(z) = - \sqrt{\tfrac{\pi}{\beta \sinh (\pi \beta)\+}\+}\+ \sin\!\big( \beta \log \tfrac{z}{2} - \vartheta_\beta \big) + O(z^2),\qquad \text{for }\, z\to 0^+\-,\label{eq:originMacdonald}
\end{gather}
\end{subequations}
with $\vartheta_\beta \vcentcolon= \arg \Gamma(1 + i \beta)\in (-\pi,\pi]$.\newline
Hence, the problem~\eqref{eq:entireProblem} has the following solution:
\begin{equation}
    u_\lambda(r)=\begin{dcases}
        A\, w_\lambda (r),\qquad &0 \leq r \leq r_0,\\
        B\+ \sqrt{r\+}\+ K_{i\beta}(\lambda\+ r),\quad & r>r_0,
    \end{dcases}
\end{equation}
with $A, B$ proper constants such that
$$\lim_{r\to r_0^-} u_\lambda(r)=\lim_{r\to r_0^+} u_\lambda(r),\qquad \lim_{r\to r_0^-} u_\lambda'(r)=\lim_{r\to r_0^+} u_\lambda'(r).$$
The previous matching condition is due to the embedding $H^2(\R)\hookrightarrow C^{\+1}(\R)$.
Therefore, one must require
\begin{equation}\label{eq:linearSystem}
    \begin{dcases}
        w_\lambda(r_0)\,A-\sqrt{r_0\+}\+K_{i\beta}(\lambda\+r_0)\+B=0,\\
        w_\lambda'(r_0)\, A-\left[\frac{1}{2\sqrt{r_0\+}\+}\+ K_{i\beta}(\lambda\,r_0)+\lambda \sqrt{r_0\+}\+K_{i\beta}'(\lambda\+r_0)\right]\!B=0.
    \end{dcases}
\end{equation}
In order to exclude the identically zero solution $u_\lambda\equiv 0$, one cannot admit $A=B=0$.
Consequently, the matrix associated with the previous linear system of equations must be singular:
$$-w_\lambda(r_0)\left[\frac{1}{2\sqrt{r_0\+}\+}\+ K_{i\beta}(\lambda\,r_0)+\lambda \sqrt{r_0\+}\+K_{i\beta}'(\lambda\+r_0)\right]+\sqrt{r_0\+}\+K_{i\beta}(\lambda\+r_0)\,w_\lambda'(r_0)=0,$$
namely,
\begin{equation}\label{eq:vanishingDeterminant}
    \left[\frac{w_\lambda(r_0)}{2}-r_0\, w_\lambda'(r_0)\right]\- K_{i\beta}(\lambda\,r_0)+\lambda\+ r_0\,w_\lambda(r_0)\, K_{i\beta}'(\lambda\+r_0)=0.
\end{equation}
Our goal is to prove that~\eqref{eq:vanishingDeterminant} can be solved for infinitely many energy levels accumulating at zero according to Efimov's asymptotic geometrical law.\newline
Combining~\cite[eqs. 10.27.3, 10.29.1, and 10.30.2]{OlMa10}, one can compute the asymptotic expansion for the derivative of the Macdonald function as $z\to 0^+$
\begin{align*}
    K_{i\beta}'(z)&=-\frac{1}{2\+z}\left[\Gamma(1\--i\beta)(\tfrac{1}{2}z)^{i\beta}+\Gamma(1\-+i\beta)(\tfrac{1}{2}z)^{-i\beta}\right]+ O(z)\\
    &=-\frac{\abs{\Gamma(1\-+i\beta)}}{2\+z}\left[e^{-i \vartheta_{\-\beta}}\+e^{i\beta \ln(\frac{1}{2}z)}+e^{i \vartheta_{\-\beta}}\+e^{-i\beta \ln(\frac{1}{2}z)}\right]+O(z),
\end{align*}
where we have used $\Gamma(\conjugate{z})=\conjugate*{\Gamma(z)}$. Hence, by exploiting~\cite[eq. 5.4.3 p. 137]{AsRo10} for the absolute value of the Gamma function, one gets
\begin{equation}\label{eq:originMacdonaldDerivative}
    z\+K_{i\beta}'(z)=-\sqrt{\tfrac{\pi \beta}{\sinh(\pi\beta)\+}\+}\+\cos\!\big(\beta\ln\tfrac{z}{2}-\vartheta_\beta\big)+O(z^2),\qquad\text{for }\, z\to 0^+.
\end{equation}
Introducing the parameters
\begin{align}\label{non-vanishingTogether}
    a\vcentcolon=\frac{w_{\+0}(r_0)}{2}-r_0\, w_{\+0}'(r_0), && b\vcentcolon=\beta \, w_{\+0}(r_0),
\end{align}
condition~\eqref{eq:vanishingDeterminant} can be equivalently rewritten as
\begin{equation}\label{eq:toBeSolved}
    -a \sin\!\big(\vartheta_\beta-\beta\+ \ln\-\tfrac{\lambda\+ r_0}{2}\big)+b\cos\!\big(\vartheta_\beta-\beta \+\ln\-\tfrac{\lambda\+ r_0}{2}\big)=F_{\mspace{-0.75mu}r_0}(\lambda),
\end{equation}
for some continuous function $F_{\mspace{-0.75mu}r_0}: \Rplus\longrightarrow \R$ satisfying
$$\lim_{\lambda\to 0^+} \!F_{\mspace{-0.75mu}r_0}(\lambda)=0, \qquad\text{for all }\,r_0\-\in\Rplus.$$
We stress that $a$ and $b$ cannot both be equal to zero.
Otherwise, it follows that $w_{\+0}(r_0)=w_{\+0}'(r_0)=0$, and the initial value problem~\eqref{eq:innerProblem} for $\lambda=0$ with such conditions has the identically zero function as a solution.
By uniqueness, $w_{\+0}(r)=0$ for all $r\in [0,r_0]$, contradicting $w_\lambda'(0)\neq 0$.\newline
Let us solve~\eqref{eq:toBeSolved} in the case where $F_{\mspace{-0.75mu}r_0}$ is replaced by zero:
\begin{align}
    -a &\sin\!\big(\vartheta_\beta-\beta\+ \ln\-\tfrac{\lambda\+ r_0}{2}\big)+b\cos\!\big(\vartheta_\beta-\beta \+\ln\-\tfrac{\lambda\+ r_0}{2}\big)=0\label{eq:homogeneousEquation}\\
    \iff &\tan\big(\vartheta_\beta-\beta\+\ln\-\tfrac{\lambda\+r_0}{2}\big)=\frac{b}{a},\qquad \text{provided }\, a\neq 0,\nonumber
    \intertext{thus,}
    &\vartheta_\beta-\beta\+\ln\-\tfrac{\lambda\+r_0}{2}=\arctan\tfrac{b}{a}+n\+\pi,\qquad n\in\Z\nonumber\\
    \implies &\lambda_n^0=\tfrac{2}{r_0}\, e^{\frac{\vartheta_{\-\beta}-\,\alpha}{\beta}}\,e^{-n\,\frac{\pi}{\beta}},\qquad \alpha\vcentcolon=\arctan{\tfrac{b}{a}}.\label{eq:homogeneousSolution}
\end{align}
Note that $\alpha=0$ in the case $b=0$, while~\eqref{eq:homogeneousSolution} remains valid for $a=0$ upon choosing $\alpha=\frac{\pi}{2}$.
In the following, we show that equation~\eqref{eq:toBeSolved} admits infinitely many solutions $\lambda_n$ with the same asymptotic behaviour as $\lambda_n^0$ as $n\to\pInfty$.\newline
To this end, let us consider the equation
\begin{equation}\label{eq:toBeSolved2}
    \sin \eta =\frac{(-1)^n}{\!\-\sqrt{a^2\-+b^2\+}\+}\+F_{\mspace{-0.75mu}r_0}\-\big(\lambda_n^0\,e^{\frac{\eta}{\beta}}\big),\qquad \eta\in \big(\!-\-\tfrac{\pi}{2},\tfrac{\pi}{2}\+\big).
\end{equation}
Since $\lambda_n^0\,e^{\frac{\eta}{\beta}}\in \big(\lambda_n^0\,e^{-\frac{\pi}{2\+\beta}},\,\lambda_n^0\,e^{\frac{\pi}{2\+\beta}}\big)$ and $\lim\limits_{n\to\pInfty}\!F_{\mspace{-0.75mu}r_0}\-\big(\lambda_n^0\,e^{\frac{\eta}{\beta}}\big)=0$, there exists $n_0=n_0(v,r_0,\beta)\in\Z$ such that
$$\frac{\abs{F_{\mspace{-0.75mu}r_0}\-\big(\lambda_n^0\,e^{\frac{\eta}{\beta}}\big)}}{\sqrt{a^2\-+b^2\+}}<1,\qquad \forall n\geq n_0.$$
Therefore, for all $n\geq n_0$,~\eqref{eq:toBeSolved2} has a unique solution $\eta_{\+n}\-\in\-\big(\!-\-\tfrac{\pi}{2},\tfrac{\pi}{2}\+\big)$ satisfying $\lim\limits_{n\to\pInfty}\eta_{\+n}=0$.\newline
We now have all the elements required to prove the following proposition.
\begin{prop}\label{pro:nonhomogeneousSolutions}
    Let $\lambda_n^0\!\in\-\Rplus$ be a solution to~\eqref{eq:homogeneousEquation}.
    Then, there exists $n_0=n_0(v,r_0,\beta)\in\Z$ such that if $n\geq n_0$, there exists a unique solution $\eta_{\+n}\!\in\-\big(\!-\-\tfrac{\pi}{2},\tfrac{\pi}{2}\+\big)$ to~\eqref{eq:toBeSolved2}, with $a,b$ given by~\eqref{non-vanishingTogether}.
    Moreover, equation~\eqref{eq:vanishingDeterminant} admits infinitely many solutions of the form
    \begin{equation}\label{eq:energyLevels}
        \lambda_n=\lambda_n^0\, e^{\,\eta_{\+n}/\beta},\qquad n\geq n_0.
    \end{equation}
    \begin{proof}
        We have already discussed the existence and uniqueness of $\eta_n$; therefore, we focus on the proof of~\eqref{eq:energyLevels}.
       Equation~\eqref{eq:vanishingDeterminant} is equivalent to~\eqref{eq:toBeSolved}, which can, in turn, be rewritten by replacing $\lambda$ with $\lambda^0_n \,e^{\frac{\eta_{\+n}}{\beta}}$
        $$-a\sin(\alpha + n\+\pi-\eta_{\+n})+b\cos(\alpha + n\+\pi-\eta_{\+n})=F_{\mspace{-0.75mu}r_0}\big(\lambda_n^0\,e^{\frac{\eta_{\+n}}{\beta}}\big).$$
        Hence,
        $$-a\+\big(\sin \alpha \+\cos \eta_{\+n} \-- \cos \alpha \+\sin \eta_{\+n}\big)+b\+\big(\cos\alpha \+\cos \eta_{\+n}\-+\sin\alpha\+\sin\eta_{\+n}\big)=(-1)^n F_{\mspace{-0.75mu}r_0}\big(\lambda_n^0\,e^{\frac{\eta_{\+n}}{\beta}}\big).$$
        Since $-a\sin\alpha+b\cos\alpha=0$ (see~\eqref{eq:homogeneousEquation}), one has
        \begin{equation}\label{intermediateStep}
            \big(a\+\cos\alpha+b\+\sin\alpha\big)\sin\eta_{\+n}=(-1)^n F_{\mspace{-0.75mu}r_0}\big(\lambda_n^0\,e^{\frac{\eta_{\+n}}{\beta}}\big).
        \end{equation}
        Lastly, exploiting the identities
        \begin{align*}
            \cos\alpha=\frac{a}{\!\-\sqrt{a^2\-+b^2}}, && \sin\alpha=\frac{b}{\!\-\sqrt{a^2\-+b^2}},
        \end{align*}
        it is straightforward that~\eqref{intermediateStep} is equivalent to~\eqref{eq:toBeSolved2} with $\eta=\eta_{\+n}$; moreover,~\eqref{eq:toBeSolved2} is automatically solved for $n\geq n_0$.
        This concludes the proof.
        
    \end{proof}
\end{prop}
\n The preceding analysis yields the following proposition.
\begin{prop}\label{pro:eigenvaluesEigenfunctions}
    Using the notation of Proposition~\ref{pro:nonhomogeneousSolutions}, the eigenvalue problem~\eqref{eq:sWaveEigenvalueProblem} %-- describing the Hamiltonian $-\frac{1}{\mu}\Delta+\mathcal{E}, H^2(\R^3)$ in its subspace of zero angular momentum -- 
    has infinitely many negative solutions $\{E_{\mathrm{BO};\,n}\}_{n\+=\+n_0}^\infty$, with $\lim\limits_{n\to\infty} E_{\mathrm{BO};\,n}=0$.
    Moreover, for all $n\geq n_0$
    $$E_{\mathrm{BO};\,n}=-\frac{4}{\mu\+r_0^2}\,e^{\frac{2}{\beta}\big(\vartheta_\beta\+-\,\arctan\frac{b}{a}\++\,\eta_n\big)}\,e^{-\frac{2\pi}{\beta}\+ n}$$
    and the corresponding eigenfunction is
    $$u_n(r)=\begin{dcases}
        A_n \,w_{\lambda_n}(r),\qquad & 0 \leq r \leq r_0,\\
        B_n \+\sqrt{r\+}\+ K_{i\beta}(\lambda_n \+r),\qquad & r>r_0
    \end{dcases}$$
where
    $$A_n=\begin{dcases}
        \frac{\sqrt{r_0\+}}{w_{\lambda_n}\-(r_0)}\,K_{i\beta}(\lambda_n\+r_0) \+B_n,\qquad & \text{if }\, w_{\lambda_n}\-(r_0)\neq 0,\\
        \frac{K_{i\beta}(\lambda_n\+r_0)+2\+\lambda_n\+ r_0 \,K_{i\beta}'(\lambda_n\+r_0)}{2\+\sqrt{r_0\+}\+w_{\lambda_n}'\!(r_0)}\,B_n,\qquad &\text{if }\,w_{\lambda_n}\-(r_0)=0,
    \end{dcases}$$
    and $B_n$ is the normalisation constant.
    \begin{proof}
        The result is derived from Proposition~\ref{pro:nonhomogeneousSolutions} and by observing that combining the assumption $u_\lambda'(0)\neq 0$ (implying $w_{\lambda}(r_0)\neq 0 \lor w_\lambda'(r_0)\neq 0$) with equation~\eqref{eq:vanishingDeterminant}, the linear system~\eqref{eq:linearSystem} has a corresponding coefficient matrix of rank $1$.

    \end{proof}
\end{prop}
\n As a consequence of the previous proposition, one obtains the proof of the asymptotic geometrical behaviour of the energy levels in the Born-Oppenheimer approximation 

$$\lim_{n \to \infty} \frac{E_{\mathrm{BO};\,n}}{E_{\mathrm{BO};\,n\++1}}=
\lim_{n \to \infty} e^{\frac{2}{\beta}(\eta_n-\,\eta_{n\++1})}\, e^{\frac{2\pi}{\beta}}= e^{2\pi/\beta}.$$

%-------------------------------
% ########  #### ########      
% ##     ##  ##  ##     ##     
% ##     ##  ##  ##     ##     
% ########   ##  ########      
% ##     ##  ##  ##     ##     
% ##     ##  ##  ##     ## ### 
% ########  #### ########  ###
%-------------------------------
\vs

\bibliographystyle{plainnat}
%\bibliography{references}

\end{document}